\documentclass[prl,amsmath,aps,twocolumn,superscriptaddress]{revtex4}
\usepackage{graphicx,xspace}
\usepackage{float}
\usepackage{amsmath}
\usepackage{amssymb}
\usepackage[normalem]{ulem}
\input{epsf}
\begin{document}
\newcommand{\scalar}[2]{\left \langle#1\ #2\right \rangle}
\newcommand{\me}{\mathrm{e}}
\newcommand{\mi}{\mathrm{i}}
\newcommand{\dif}{\mathrm{d}}
\newcommand{\period}{\text{per}}
\newcommand{\free}{\text{fr}}
\newcommand{\eq}[1]{Eq.~(\ref{e:#1})}
\newcommand{\eqq}[1]{Eq.~(\ref{e:#1})}
\newcommand{\EQ}[1]{(\ref{e:#1})}
\newcommand{\eqtwo}[2]{equations~(\ref{e:#1}) and~(\ref{e:#2})}
\newcommand{\EQTWO}[2]{Equations~(\ref{e:#1}) and~(\ref{e:#2})}
\newcommand{\fig}[1]{Fig.~\ref{f:#1}}
\newcommand{\FIG}[1]{Fig.~\ref{f:#1}}
\newcommand{\quot}[1]{\lq#1\rq}
\newcommand{\eg}{\textrm{e.g.}}
\newcommand{\cf}{\textrm{cf}}
\newcommand{\etc}{\textrm{etc}}
\newcommand{\ie}{\textrm{i.e.}}
\newcommand{\SET}[1]{\{#1\}}
\newcommand{\expl}[1]{\exp \left[ #1 \right] } 
\newcommand{\lb}{\left[}  
\newcommand{\rb}{\right]}  
\newcommand{\lc}{\left(}  
\newcommand{\rc}{\right)}  
\newcommand{\mult}{\times} 
\newcommand{\multcc}{\cdot} 
\newcommand{\multnn}{\cdot} 
\newcommand{\ran}{\sub{ran}}
\newcommand{\dd}[1]{\text{d}{#1\ }}   
\newcommand{\ddd}[1]{\text{d}{#1}}   
\newcommand{\scal}[2]{(#1 \pmb{\cdot} #2)}
\newcommand{\mean}[1]{\left\langle #1 \right\rangle}
\newcommand{\half}{\frac{1}{2}}

\title{Geometry of Gaussian signals} 
\author{Alberto Rosso}
\email{rosso@lptms.u-psud.fr}
\affiliation{Laboratoire de Physique Th\'{e}orique et Mod\`{e}les Statistiques \\
B\^{a}t. 100, Universit\'{e} Paris-Sud, 
91405 Orsay Cedex, France}
\author{Raoul Santachiara}
\email{rsantach@science.uva.nl}
\affiliation{Instituut voor Theoretische Fysica\\
Valckenierstraat 65, 1018 XE Amsterdam,The Netherlands}
\author{Werner Krauth}
\email{krauth@lps.ens.fr}
\affiliation{CNRS-Laboratoire de Physique Statistique\\
Ecole Normale Sup{\'{e}}rieure, 
24 rue Lhomond, 75231 Paris Cedex 05, France}
\begin{abstract} 
We consider Gaussian signals, \ie\ random functions $u(t)$  ($t/L \in
[0,1]$) with independent Gaussian Fourier modes of variance $\sim
1/q^{\alpha}$, and compute their statistical properties in small
windows $[x, x+\delta]$. We determine moments of the probability
distribution of the mean square width of $u(t)$ in powers of the window
size $\delta$.  We show that the moments,  in the small-window limit
$\delta \ll 1$, become universal, whereas they strongly depend on the
boundary conditions of $u(t)$ for larger $\delta$.  For $\alpha > 3$,
the probability distribution is computed in the small-window limit and
shown to be independent of $\alpha$.
\end{abstract}
\maketitle
Gaussian signals---random functions with independent Gaussian
Fourier components of variance $\sim 1/q^{\alpha}$ ---have been used to
describe physical situations ranging from $1/f$ noise in electric
circuits \cite{surya} to intermittency in turbulent flows \cite{frisch}, and
to interfaces in random media \cite{kardar,barabasi_book}.

The exponent $\alpha > 1$ of the power spectrum 
fixes the signal's mean square width $w_2(L)$ on a scale $L$, as
characterized by the roughness exponent $\zeta$
\begin{equation}
w_2(L) \equiv \frac{1}{L} \int_0^L \dd{t} u(t)^2 - \lb \frac{1}{L} 
\int_0^L \dd{t} u(t) \rb ^2 \sim L^{2 \zeta}, 
\label{e:power_zeta}
\end{equation}
which, in one dimension, is given by $\zeta = \half(\alpha -1)$.
The notorious random walk, with roughness $\zeta = \half$, corresponds
to a power spectrum with $\alpha=2$. The well-studied  curvature-driven
model \cite{plischke.curvature.94} also belongs to this class of systems
and corresponds to the case $\alpha=4$.

For elastic interfaces in disordered media and other systems with non-integer
$\alpha$, this exponent of the power spectrum and equivalently the roughness
exponent are extremely difficult to calculate \cite{nattermann_stepanow_depinning,narayan,chauve,rosso_krauth_manifold}.
Exact results are rare \cite{kardar2}.  In contrast to the random walk and the
curvature-driven model, these systems are not exactly Gaussian.  This means
that the Fourier modes, even at small $q$, are correlated. More intricate
statistical properties than the scaling of $w_2$ are able to expose these
correlations. A common approach consists in studying the probability
distribution of the mean squared width $w_2$, which fluctuates from sample to
sample \cite{racz.random.94,bramwell.xy,rosso_width}.  This distribution has
been used to characterize the geometric properties of numerical and
experimental data \cite{marinari_racz,rosso_width2,vandembroucqphi,zapperi}.

Non-Gaussian corrections for the probability distribution of the
mean-square width were explicitly determined in a non-trivial depinning
problem and found to be on the $0.1 \%$ level \cite{rosso_width}. It
was possible to understand this small effect because non-Gaussian
corrections appear only in high orders of perturbation theory
\cite{rosso_width,ledoussal_Rcalcul}. The excellent agreement between
complicated physical models on the one side and their effective Gaussian
description on the other motivates a finer analysis of the universal
statistical properties of Gaussian signals, which is the object of
this letter.

\begin{figure}
   \centerline{\includegraphics{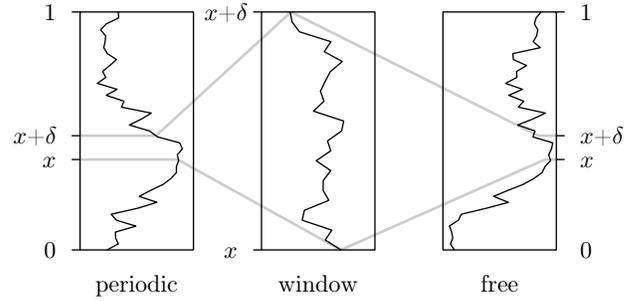}}
   \caption{Periodic (left) and free (right) Gaussian signals \mbox{$u(t\multcc L)$}. 
   This paper studies the statistics of signals inside a small window (middle).}   
\label{f:three_models}
\end{figure}

Because of their definition in Fourier space, it is most convenient to
study periodic signals with $u(t)= u(t+L)$.  However, experimental
systems are usually non-periodic. \emph{Free} boundary conditions are
commonly modeled by Gaussian signals $u(t)$  with zero mean and vanishing
derivatives at the end-points (see \fig{three_models}).
The probability distribution of $w_2$ depends on the boundary conditions:
The different distributions of free and periodic signals were computed
analytically \cite{antal.1overf.01,rosso_width2}.

Several authors have studied the signal $u(t)$ inside a small window
(see \fig{three_models}), \ie\ the piece  with $t/L \in [x,x +
\delta]$. Antal et al. \cite{antal.1overf.01} performed numerical
simulation for different values of $\alpha$. It was empirically found
that the probability distribution of the mean squared width inside a
window agrees well with free boundary conditions for $\alpha \simeq 2$
\cite{rosso_width2}. The two distribution differ markedly  for $\alpha$
outside this range, as was clearly shown by de Queiroz \cite{queiroz}.

In this paper, we calculate analytically the statistical properties
of Gaussian signals, with periodic or free boundary conditions, in
small windows: We determine the expansion in powers of the window size
$\delta$ of the lower moments of the probability distribution of $w_2$
and find them to be identical for the periodic and for the free signal far
from the boundary. This is important for the analysis of experimental
data, for which the boundary conditions can usually not be controlled:
the statistics in a small window is independent of them, \ie\ contains
only universal information. We complement this determination of moments
by numerical calculations of the probability distribution itself.

The Gaussian signal is defined by the action
\begin{equation}
 S= \frac{1}{2}\int_0^L \dd{t} \lb \frac{\partial^{\alpha/2}
 u(t)}{\partial t^{\alpha/2}} \rb^2, 
\label{e:action}
\end{equation}
where the derivative for non-integer $\alpha$ is understood in Fourier space
\cite{antal.1overf.01, LongPaper}.  For a periodic signal of zero mean, 
\begin{equation}
u(t)= \sum_{n=1}^{\infty} a_n \cos \lb \frac{2 \pi n}{L} t \rb +
b_n \sin\lb \frac{2 \pi n}{L} t \rb,   
\label{e:Fourier_sine}
\end{equation}
this means that the $a_n$ and $b_n$ are Gaussian random numbers
of variance $\sigma_n^2= L^{2 \zeta} 2^{- 2 \zeta} \multcc (\pi
n)^{-\alpha}$.  \emph{Free} Gaussian signals are commonly modeled by
cosines with period $2L$, 
\begin{equation}
u(t)= \sum_{n=1}^{\infty} c_n \cos\left(\frac{\pi n}{L}t \right), 
\label{e:Fourier_cosine}
\end{equation}
where the $c_n$ are of variance $2 L^{2 \zeta} \multcc (\pi n)^{-\alpha}$.

The mean squared width of a signal $u(t\multcc L)$ in a window 
$ [x,x+\delta]$ is 
\begin{equation*}
w_2(x,\delta)= \frac{1}{\delta}\int_x^{x+\delta} 
u^2(t\multcc L) dt -\left(\frac{1}{\delta}
\int_x^{x+\delta} u(t \multcc L) \dd{t} \right)^2.
\end{equation*}
For the free boundary conditions, we have
\begin{gather}
w_2^{\free}(x,\delta) = \sum_{n,m=1}^{\infty} c_n c_m D_{nm}(x, \delta)
\label{e:w_2_equation}
\end{gather}
with
\begin{multline*}
D_{nm}(x, \delta)= \frac{1}{\delta} \int^{\delta+x}_x \cos( \pi n t) \cos( \pi m t) \ddd{t} \\
-\frac{1}{\delta^2} \int^{\delta+x}_x \cos( \pi n t) \ddd{t} \int^{ \delta+x}_x
\cos(\pi m t) \ddd{t}.
\label{e:Fourier_width_free}
\end{multline*}
The periodic signal's mean squared width, $w_2^{\period}$, has an
analogous expression in terms of cos-cos, cos-sin and sin-sin integrals.
Clearly, $w_2^{\period}$ does not depend on the initial point $x$, but only 
on the window size $\delta$.

The above equations allow to compute $w_2$ for one given sample (choice
of $\SET{a_n,b_n}$ or $\SET{c_n}$) and its probability distribution,
which is characterized by the ensemble average $\mean{w_2}$
and by the rescaled distribution $\phi(z)$ with $z = w_2/\mean{w_2}$.
For free boundary conditions, \eq{w_2_equation} implies
\begin{equation}
\mean{w_2^{\free}(x,\delta)} = \frac{2 L^{2 \zeta}}{\pi^{\alpha}}  \sum_{n=1}^{\infty}
\frac{D_{nn}(x,\delta)} {n^{\alpha}}.
\label{e:D_nn_sum}
\end{equation}
This gives for the variance of the rescaled probability distribution 
$\phi(z)$
\begin{equation}
\kappa_2^{\free}(x,\delta)= \mean{\frac{2 \sum_{n,m=1}^{\infty} \frac{D^2_{nm}(x,\delta)}{n^{\alpha}
m^{\alpha}}} { \lb \sum_{n=1}^{\infty} \frac{D_{nn}(x,\delta)}{n^{\alpha}}\rb^2}}, 
\label{e:variance_general}
\end{equation} 
which is independent of $L$. All the other cumulants of $\phi(z)$ are 
scale-free, and are defined analogously through multiple sums. 
We note that for $\delta=1$, and with the normalization \EQ{action}, we have
\begin{gather*}
\mean{w_2^{\free}(\delta=1) }= \frac{L^{2 \zeta}}{ \pi^{1+ 2 \zeta}} \zeta(\alpha)\\
\mean{w_2^{\period}(\delta=1)}= \frac{L^{2 \zeta}}{2^{2 \zeta} \pi^{1+ 2 \zeta}} \zeta(\alpha), 
\end{gather*} 
where $\zeta(\alpha)$ is the Riemann zeta function.
All the (scale-free) cumulants are known for $\delta=1$  \cite{antal.1overf.01,rosso_width2}
\begin{gather}
\kappa_n^{\free}(\delta=1) = (2n-2)!!\frac{\zeta(n \alpha)}{ \zeta^n(\alpha)}\\
\kappa_n^{\period}(\delta=1) =  (n-1)!\frac{\zeta(n \alpha)}{\zeta^n(\alpha)}.
\label{e:cumulant_free_period}
\end{gather} 

Sums as in \EQTWO{D_nn_sum}{variance_general} may be evaluated with
a powerful formula \cite{LongPaper}
\begin{equation}
\begin{split}
\sum_{n=1}^{\infty} \frac{f(n\delta)}{n^{\alpha}} = \delta^{\alpha -1} 
\int_{0}^{\infty} \dd{t} \lb \sum_{m= \lfloor \alpha \rfloor}^{\infty} 
\frac{f^m(0) t^{m-\alpha}}{m!} \rb \\+ \sum_{m=0}^{\infty} \delta^m f^m(0)
\frac{\zeta(\alpha - m)}{m!}
\end{split}, 
\label{e:sum_integral_zeta}
\end{equation}
where $\lfloor \alpha\rfloor$ is the integer part of $\alpha$. The
\eq{sum_integral_zeta}, which is in the spirit of the Euler-Maclaurin
formula, is valid only for analytic functions $f$ and non-integer
$\alpha$: the first term on the  right can be interpreted as the naive
limit of the sum as $\delta \rightarrow 0$, with $t=n\delta$. The second
term on the right contains the Taylor expansion of $f(n\delta)$ around
zero. For integer $\alpha$, the singularity of $\zeta(1)$ generates
additional logarithms, which are explicitly known \cite{LongPaper}.

We use \eq{sum_integral_zeta}, and its generalization for integer
$\alpha$,  to compute moments of the probability distribution of $w_2$ in
a window $[x,x+ \delta]$ (for a periodic signal, the result is
independent of $x$, for a free one, this is true only for  $\alpha = 2$).
For periodic boundary conditions, we find
\begin{multline}
\frac{\mean{w_2^{\period}(\delta)}}{L^{2 \zeta}} = 
\frac{2^{-\alpha-1}}{\zeta(-\alpha-1)} 
\frac{\zeta(\alpha+2)}{\pi^{\alpha+2}} \delta^{\alpha-1} + \\
\frac{2^{1-\alpha}}{3} \frac{\zeta(\alpha-2)}{ \pi^{\alpha-2}} \delta^2-
\frac{ 2^{2-\alpha}}{45} \frac{\zeta(\alpha-4)}{ \pi^{\alpha-4}} \delta^4 + \cdots
\label{e:w2_series}
\end{multline}
This formula gets modified by logarithms for $\alpha=3$, $\alpha=5$,
etc \cite{LongPaper}. 
Interestingly, an analogous expansion appears in the correlation
function governing the density of zero-crossings of a Gaussian
signal \cite{Majumdar}.
We note that, for even  $\alpha$, the series \EQ{w2_series}
stops, because the Riemann zeta function vanishes for even negative integers.
For example, the periodic random walk ($\alpha=2$) and the driven curvature
model ($\alpha=4$) yield 
\begin{equation}
\frac{\mean{w_2^{\period}(\delta)} }{L^{2 \zeta}}= 
\begin{cases}
\frac{\delta}{6} - \frac{\delta^2}{12}\ &  (\alpha=2)\\
\frac{\delta^2}{144}- \frac{\delta^3}{120} +\frac{\delta^4}{360}\ & (\alpha=4)
\end{cases}.
\label{e:}
\end{equation}

From the series \EQ{w2_series}, the dominant term of
$\mean{w_2^{\period}(\delta)}$ scales for small windows as
$(L \delta) ^{\alpha-1} = (L \delta)^{2 \zeta}$, in agreement with
the self-affinity relation \EQ{power_zeta}.  However, this natural
scaling breaks down for $\alpha > 3$, \ie\ for roughness exponents
$\zeta>1$. There, the small window scaling, from \eq{w2_series}, is as
$(L\delta)^2 \multcc L^{2\zeta-2}$. In addition to an $\alpha$-independent
scaling at small distances, a scale factor, depending on $\alpha$
and on the system size appears explicitly.  This was pointed out by
Tang and Leschhorn \cite{leschhorn_tang} in the context of depinning,
where models with $\zeta > 1$ ($\alpha > 3$) appear naturally
\cite{jensen,rosso_krauth_non-harmonique}.

For $\alpha < 3$, and free  boundary conditions with $0<x<1$, the dominant
integral in \eq{sum_integral_zeta} involves an intricate double limit, where $\delta
\rightarrow 0$ and $x/\delta \rightarrow \infty$.  Rapidly oscillating
terms with vanishing contributions, generated from the second limit,
need to be discarded. We find \cite{LongPaper} that the dominant term of
$\mean{w_2^{\free}}$  proportional to $\delta^{\alpha-1}$ is identical
to the dominant term of $\mean{w_2^{\period}}$, from \eq{w2_series}. The
mean squared width $\mean{w_2} $ is thus insensitive to the boundary
conditions. We note that the expansion of $\mean{w_2}$ obtained from
\eq{sum_integral_zeta} provides non-intuitive explicit prescriptions for
extracting the roughness exponent from experimental or numerical data
in powers of $\delta$, which differ from the standard ansatz \cite{jensen}.

For $\alpha>3$, the  dominant term for small $\delta$ is
\begin{equation}
\frac{\mean{w_2^{\free}(\delta)}}{L^{2\zeta}}=
\sum_{n=1}^{\infty}\frac{\sin(n\pi x)^2}{6\pi^{\alpha-2}n^{\alpha-2}} \delta^2
+ O(\delta^{\alpha-1}, \delta^4).
\label{e:}
\end{equation}
The vanishing derivatives at the end-points force $\mean{w_2^{\free}}
$ to be smaller than $\mean{w_2^{\period}}$ for all $0<x<1$.  At the
end-points $x=0$ and $x=1$, the $\delta^2$ term vanishes \cite{LongPaper}.

We now turn to the higher moments  of the distribution $\phi(z)$,
(cf. \eq{variance_general} for free boundary conditions), which can
be evaluated with straightforward generalizations of 
\eq{sum_integral_zeta} to multiple sums. The leading term of
$\kappa_n$, for $\alpha<3$, is given by an $n$--dimensional, well-behaved,
integral, which can be computed numerically.  The calculation of the
variance $\kappa_2$ is along the lines of the above determination
of $\mean{w_2}$. Discarding again rapidly oscillating terms in the
limit $\delta\rightarrow 0 $ and $x/\delta \rightarrow \infty$, we can
analytically prove \cite{LongPaper} that $\kappa_2^{\period}(\delta)$
coincides with $\kappa_2^{\free}(x, \delta)$  for $1 < \alpha
< 3$. Furthermore, in the special case $\alpha=2$, the integral
explicitly agrees with the variance $\kappa_2^{\free}(\delta=1)$  from
\eq{cumulant_free_period}, which for the random walk is independent of
$\delta$. This is due to the Markov-chain property of the free random
walk.  The density matrix generating the path is infinitely divisible,
and essentially reproduces itself for all values of $\delta$. In contrast,  
for all $\alpha \ne 2$, we  have
\begin{equation}
\kappa_2^{\period}(\delta \rightarrow 0) = 
\kappa_2^{\free}(\delta \rightarrow 0)  \ne \kappa_2^{\free}(\delta=1).
\label{e:}
\end{equation}
\begin{figure}
   \centerline{\includegraphics{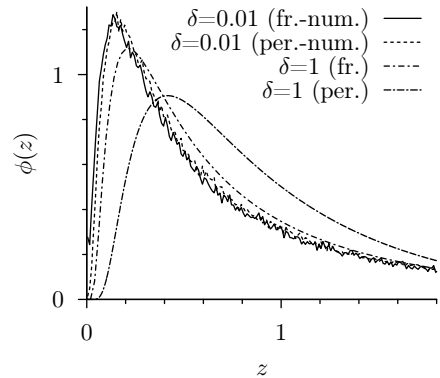}}
   \caption{Probability distribution $\phi(z)$ for free and periodic
   signals at $\alpha=2.5$.  Direct simulations for $\delta=0.01$ (with
   $x=0.495$ for the free case) are compared to analytic solutions for $
   \delta=1$ \cite{antal.1overf.01,rosso_width2}.}
\label{f:zeta_0.75_figure} \end{figure}

\quot{Windows} of periodic or free signal are thus different from
the free signal itself.  To stress this crucial result, we show in
\fig{zeta_0.75_figure} the numerically computed distribution $\phi(z)$
for $\delta=0.01$ for both free and periodic boundary conditions, and
compare them to the distributions for $\delta=1$.  The two distribution
at $\delta=0.01$ agree very well, but, for this value of $\zeta$ close
to $\half$, they differ somewhat from the free $\delta=1$ distribution.
This point was missed in previous work \cite{rosso_width2}, because the
exponent considered there was very close to $\half$.  Notice that the
distribution for the periodic $\delta=1$ signal is very different from 
the other distributions.

The calculation, for $\alpha <3 $, of cumulants $\kappa_n$ with
$n>2$ presents no conceptual difficulties, and we conjecture that all
moments, and thus the distribution itself, in the small-window limit,
are independent of boundary conditions. This is also supported by the
numerical results of \fig{zeta_0.75_figure}.

\begin{figure}
   \centerline{\includegraphics{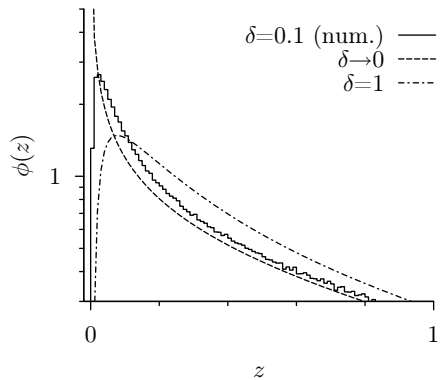}}
   \caption{Probability distribution $\phi(z)$ for free signals at 
   $\alpha=3.5$.  A direct simulation of
    \eq{w_2_equation} ($x=0.45$,  $\delta=0.1$)  with $10^6$ samples is
   compared to the function \EQ{limiting_function} in the small-window
   limit and to the analytic solution for $\delta=1$ \cite{rosso_width2}.}
\label{f:alpha_3.5_figure} \end{figure}

In the case $\alpha \ge 3$, the integral term in \eq{sum_integral_zeta}
is sub-dominant and the calculation of all cumulants becomes feasible.
We find for the moments
\begin{equation}
\kappa_n(\delta \rightarrow 0) = \begin{cases}
            (2n-2)!! + O(\delta^{\alpha -3}) & \text{for}\ \alpha > 3\\
            (2n-2)!! + O(\log \delta)       & \text{for}\ \alpha = 3
           \end{cases}.
\label{e:}
\end{equation}
This implies that the distribution $\phi(z)$ in the small-window limit
is explicitly given by
\begin{equation}
\phi(z)  = \frac{\expl{- z/2}}{\sqrt{2 \pi z}}. 
\label{e:limiting_function}
\end{equation}
The exact small-window distribution of the mean-square distribution, valid
for all $\alpha \ge 3$, was already known to be exact in the limit $\alpha
\rightarrow \infty$ \cite{antal.1overf.01}.  In \fig{alpha_3.5_figure},
the small-window distribution is compared to the numerically obtained
histogram for $\alpha=3.5$. Notice that, for $\alpha>3$, a small window
of the free signal has completely different statistics from the free
signal itself.

In conclusion, we considered in this paper statistical properties of
Gaussian signals. We studied the influence of   boundary conditions
on the signal in a small window.  An exact sum formula, non-trivial
generalization of the Euler-Maclaurin equation, allowed to reach a
systematic procedure for computing moments of the mean square distribution
function which, at small $\delta$, was found to be independent of the
boundary condition.

The expansion  in powers of the window size $\delta$  changes at
$\alpha=3$ (corresponding to a roughness of $\zeta=1$). Above this value,
the calculation of all moments of the distribution function (in the
$\delta \rightarrow 0$ limit) becomes particularly simple, and the whole
probability distribution was computed.  Clearly, the small-window limit
studied in this paper plays an important role: It is independent of
the boundary conditions and contains the true universal information  a
Gaussian signal.

\section{acknowledgment}
We thank C. Texier and J. Bouttier for helpful discussions. R.~S. thanks LPTMS
in Orsay for hospitality for a part of this work.


\end{document}